\documentclass[12pt]{article}
\usepackage[utf8]{inputenc}
\usepackage[english]{babel}
\usepackage{latexsym,amsmath,amssymb,color}
\usepackage{amsmath}
\usepackage{amsfonts}
\usepackage{amsthm}
\usepackage{amssymb}
\usepackage{stmaryrd}
\usepackage{url}
\usepackage{bbold}
\usepackage[T1]{fontenc}
\usepackage{lmodern}
\usepackage[pdftex]{graphicx}
\usepackage{float}

\newcommand{\dW}{\ensuremath{\mathbb{W}}}

\newcommand{\cW}{\ensuremath{\mathcal{W}}}

\begin{document}

\begin{center}
\section*{Compound Hawkes Processes in Limit Order Books}
{\sc Anatoliy Swishchuk}\\University of Calgary, University Drive NW, Calgary, Canada T2N 1N4\\ 
\vspace{0.3cm}
{\sc Bruno Remillard}\\HEC, 3000, chemin de la Cote-Sainte-Catherine,
Montr\'{e}al, Qu\'{e}bec, 
Canada H3T 2A7\\
\vspace{0.3cm}
{\sc Robert Elliott}\\University of Calgary, University Drive NW, Calgary, Canada T2N 1N4 \& School of Commerce, University of South Australia, Adelaide, Australia\\ 

\vspace{0.3cm}

{\sc Jonathan Chavez-Casillas}\\University of Calgary, University Drive NW, Calgary, Canada T2N 1N4\\

\end{center}

\hspace{1cm}

{\bf Abstract:} In this paper we introduce two new Hawkes processes, namely, compound and regime-switching compound Hawkes processes, to model the price processes in limit order books. We prove Law of Large Numbers and Functional Central Limit Theorems (FCLT) for both processes. The two FCLTs are applied to limit order books where we use these asymptotic methods to study the link between price volatility and order flow in our two models by using the diffusion limits of these price processes. The volatilities of price changes are expressed in terms of parameters describing the arrival rates and price changes. We also present some numerical examples.

\vspace{0.5cm}

{\bf Keywords}: Hawkes process; compound Hawkes process; regime-switching compound Hawkes processes; limit order books; diffusion limits; Law of Large Numbers;

\section{Introduction}

The Hawkes process (HP) is named after its creator Alan Hawkes [1971, 1974]. The HP is a so-called "self-exciting point process" which means that it is a point process with a stochastic intensity which, through its dependence on the history of the process, captures the temporal and cross sectional dependence of the event arrival process as well as the 'self-exciting' property observed in empirical analysis. HPs have been used for many applications, such as modelling neural activity, genetics [Carstensen, 2010], occurrence of crime [Louie et al., 2010], bank defaults and earthquakes.

The most recent application of HPs is in financial analysis, in particular, to model limit order books, (e.g., high frequency data on price changes or arrival times of trades). In this paper we study two new Hawkes processes, namely, compound and regime-switching compound Hawkes processes to model the price processes in the limit order books. We prove a Law of Large Numbers and Functional Central Limit Theorems (FCLT) for both processes. The latter two FCLTs are applied to limit order books where we use these asymptotic methods to study the link between price volatility and order flow in our two models by using the diffusion limits of these price processes. The volatilities of price changes are expressed in terms of parameters describing the arrival rates and price changes.  The general compound Hawkes process was first introduced in [Swishchuk, 2017] to model a risk process in insurance.

[Bowsher, 2007] was the first who applied a HP to financial data modelling. Cartea et al. [2011] applied an HP to model market order arrivals. Filimonov and Sornette [2012] and Filimonov et al. [2013] apply a HP to estimate the percentage of price changes caused by endogenous self-generated activity, rather than the exogenous impact of news or novel information. Bauwens and Hautsch [2009] use a 5-D HP to estimate multivariate volatility, between five stocks, based on price intensities.  We note, that Br\'{e}maud et al. [1996]  generalized the HP to its nonlinear form. Also, a functional central limit theorem for the nonlinear Hawkes process was obtained in [Zhu, 2013]. The 'Hawkes diffusion model' was introduced in [Ait-Sahalia {\it et al.} , 2010] in an attempt to extend previous models of stock prices and include financial contagion. Chavez-Demoulin et al. [2012] used Hawkes processes to model high-frequency financial data. Some applications of Hawkes processes to financial data are also given in [Embrechts {\it et al.}, [2011].

Cohen et al. [2014] derived an explicit filter for Markov modulated Hawkes process. Vinkovskaya [2014]  considered a regime-switching Hawkes process to model its dependency on the bid-ask spread in limit order books.  Regime-switching models for the pricing of European and American options were considered in [Buffington and Elliott, 2000)] and [Buffington and Elliott, 2002], respectively. A semi-Markov process was applied to limit order books in [Swishchuk and Vadori, 2017] to model the mid-price. We note, that a level-1 limit order books with time dependent arrival rates $\lambda(t)$ were studied in [Chavez-Casillas {\it et al.}, 2017], including the asymptotic distribution of the price process. General semi-Markovian models for limit order books were considered in [Swishchuk {\it et al.}, 2017].

The paper by Bacry {\it et al.} [2015)] proposes an overview of the recent academic literature devoted to the applications of Hawkes processes in finance. The book by Cartea {\it et al.} [2015] develops models for algorithmic trading in contexts such as executing large orders, market making, trading pairs or collecting of assets, and executing in dark pool. That book also contains link to a website from which many datasets from several sources can be downloaded, and MATLAB code to assist in experimentation with the data. A detailed description of the mathematical theory of Hawkes processes is given in [Liniger, 2009]. The paper by Laub {\it et al.} [2015] provides a background, introduces the field and historical developments, and touches upon all major aspects of Hawkes processes.

This paper is organized as follows. Section 2  gives the definitions of a Hawkes process (HP), definitions of compound Hawkes process (CHP) and regime-switching compound Hawkes process (RSCHP). These definitions are new ones from the following point of view: summands associated in a Markov chain but not are  i.i.d.r.v. Section 3 contains Law of Large Numbers and diffusion limits for CHP and RSCHP. Numerical examples are presented in Section 4.

%%%%%%%%%%%%%%%%%%%%%%%%%%%%%%%%%%%%%%%%%%%%%%%%%%%%%
\section{Definitions of a Hawkes Process (HP), Compound Hawkes Process (CHP) and Regime-switching Compound Hawkes Process \\(RSCHP)}

In this section we give definitions of one-dimensional, compound and regime-switching compound Hawkes processes. Some properties of Hawkes process can be found in the existing literature. (See, e.g., [Hawkes, 1971] and [Hawkes and Oakes, 1974], [Embrechts {\it et al.}, 2011],  [Zheng {\it et al.}, 2014], to name a few). However, the notions of compound and regime-switching compound Hawkes processes are new.

\subsection{One-dimensional Hawkes Process}

{\bf Definition 1 (Counting Process)}. A counting process is a stochastic process $N(t), t\geq 0,$ taking positive integer  values and satisfying: $N(0)=0.$ It is almost surely finite, and is a right-continuous step function with increments of size $+1.$

Denote by ${\cal F}^N(t), t\geq 0,$ the history of the arrivals up to time $t,$ that is, $\{{\cal F}^N(t), t\geq 0\},$ is a filtration, (an increasing sequence of $\sigma$-algebras).

A counting process $N(t)$ can be interpreted as a cumulative count of the number of arrivals into a system up to the current time $t.$ The counting process can also be characterized by the sequence of random arrival times $(T_1,T_2,...)$ at which the counting process $N(t)$ has jumped. The process defined by these arrival times is called a point process (see [Daley and Vere-Jones, 1988]).

{\bf Definition 2 (Point Process).} If a sequence of random variables $(T_1,T_2,...),$ taking values in $[0,+\infty),$ has $P(0\leq T_1\leq T_2\leq...)=1,$ and the number of points in a bounded region is almost surely finite, then, $(T_1,T_2,...)$ is called a point process.

{\bf Definition 3 (Conditional Intensity Function).} Consider a counting process $N(t)$ with associated histories ${\cal F}^N(t), t\geq 0.$ If a non-negative function $\lambda(t)$ exists such that
$$
\lambda(t)=\lim_{h\to 0}\frac{E[N(t+h)-N(t)|{\cal F}^N(t)]}{h},
\eqno{(1)}
$$
then it is called the conditional intensity function of $N(t)$ (see [Laub {\it et al.}, 2015]). We note, that sometimes this function is called the hazard function (see [Cox, 1955]).

{\bf Definition 4 (One-dimensional Hawkes Process)}. The \\one-dimensional Hawkes process (see [Hawkes, 1971] and [Hawkes and Oakes, 1974]) is a point process $N(t)$ which is characterized by its intensity $\lambda(t)$ with respect to its natural filtration:
$$
\lambda (t)=\lambda+\int_{0}^t\mu(t-s)dN(s),
\eqno{(2)}
$$
where $\lambda>0,$ and the response function $\mu(t)$ is a positive function and satisfies $\int_0^{+\infty}\mu(s)ds<1.$

The constant $\lambda$ is called the background intensity and the function $\mu(t)$ is sometimes also called the excitation function. We suppose that $\mu(t)\not= 0$ to avoid the trivial case, which is, a homogeneous Poisson process. Thus, the Hawkes process is a non-Markovian extension of the Poisson process.

With respect to definitions of $\lambda(t)$ in (1) and $N(t)$ (2), it follows that
$$
P(N(t+h)-N(t)=m|{\cal F}^N(t))=\left\{
\begin{array}{rcl}
\lambda(t)h+o(h), &&m=1\\
o(h),&&m>1\\
1-\lambda(t)h+o(h),&&m=0.\\
\end{array}
\right.
$$

The interpretation of equation (2) is that the events occur according to an intensity with a background intensity $\lambda$ which increases by $\mu(0)$ at each new event then decays back to the background intensity value according to the function $\mu(t).$ Choosing $\mu(0)>0$ leads to a jolt in the intensity at each new event, and this feature is often called a self-exciting feature, in other words, because an arrival causes the conditional intensity function $\lambda(t)$ in (1)-(2) to increase then the process is said to be self-exciting.

We should mention that the conditional intensity function $\lambda(t)$ in (1)-(2) can be associated with the compensator $\Lambda(t)$ of the counting process $N(t),$ that is:
$$
\Lambda(t)=\int_0^t\lambda(s)ds.
\eqno{(3)}
$$

Thus, $\Lambda(t)$ is the unique ${\cal F}^N(t), t\geq 0,$ predictable function, with $\Lambda(0)=0,$ and is non-decreasing, such that
$$
N(t)=M(t)+\Lambda(t)\quad a.s.,
$$
where $M(t)$ is an ${\cal F}^N(t), t\geq 0,$ local martingale (This is the Doob-Meyer decomposition  of $N.$)

A common choice for the function $\mu(t)$ in (2) is one of exponential decay (see \cite{H1}):
$$
\mu(t)=\alpha e^{-\beta t},
\eqno{(4)}
$$
with parameters $\alpha,\beta>0.$ In this case the Hawkes process is called the Hawkes process with exponentially decaying intensity.

Thus, the equation (2) becomes
$$
\lambda (t)=\lambda+\int_{0}^t\alpha e^{-\beta (t-s)}dN(s),
\eqno{(5)}
$$
We note, that in the case of (4), the process $(N(t),\lambda(t))$ is a continuous-time Markov process, which is not the case for the choice (2).

With some initial condition $\lambda(0)=\lambda_0,$ the conditional density $\lambda(t)$ in (5) with the exponential decay in (4) satisfies the following stochastic differential equation (SDE):
$$
d\lambda(t)=\beta(\lambda-\lambda(t))dt+\alpha dN(t), \quad t\geq 0,
$$
which can be solved (using stochastic calculus) as
$$
\lambda (t)=e^{-\beta t}(\lambda_0-\lambda)+\lambda+\int_{0}^t\alpha e^{-\beta (t-s)}dN(s),
$$
which is an extension of (5).

Another choice for $\mu(t)$ is a power law function:
$$
\lambda (t)=\lambda+\int_{0}^t\frac{k}{(c+(t-s))^p}dN(s)
\eqno{(6)}
$$
for some positive parameters $c,k,p.$ This power law form for $\lambda(t)$ in (6) was applied in the geological model called Omori's law, and used to predict the rate of aftershocks caused by an earthquake.

{\bf Remark 1}. Many generalizations  of Hawkes processes have been proposed. They include, in particular, multi-dimensional Hawkes processes [Embrechts {\it et al.}, 2011], non-linear Hawkes processes [Zheng {\it et al.}, 2014], mixed diffusion-Hawkes models [Errais {\it et al.}, 2010], Hawkes models with shot noise exogenous events [Dassios and Zhao, 2011], Hawkes processes with generation dependent kernels [Mehdad and Zhu, 2011].

\subsection{Compound Hawkes Process (CHP)}

In this section we give definitions of compound Hawkes process (CHP) and regime-switching compound Hawkes process (RSCHP). These definitions are new ones from the following point of view: summands are not i.i.d.r.v., as in classical compound Poisson process, but associated in a Markov chain.

{\bf Definition 5 (Compound Hawkes Process (CHP)).} Let $N(t)$ be a one-dimensional Hawkes process defined as above. Let also $X_t$ be ergodic continuous-time finite state Markov chain, independent of $N(t),$ with space state $X.$ We write $\tau_k$ for jump times of $N(t)$ and $X_k:=X_{\tau_k}.$ The compound Hawkes process is defined as
$$
S_t=S_0+\sum_{k=1}^{N(t)}X_k.
\eqno{(10)}
$$

{\bf Remark 2.} If we take $X_k$ as i.i.d.r.v. and $N(t)$ as a standard Poisson process in (10) ($\mu(t)=0$), then $S_t$ is a compound Poisson process. Thus, the name of $S_t$ in (10)-{\it compound Hawkes process}.

{\bf Remark 3. (Limit Order Books: Fixed Tick, Two-values Price Change, Independent Orders)}. If Instead of Markov chain we take the sequence of i.i.d.r.v. $X_k,$ then (10) becomes
$$
S_t=S_0+\sum_{i=1}^{N(t)}X_k.
\eqno{(11)}
$$
In the case of Poisson process $N(t)$ ($\mu(t)=0$) this model was used in[Cont and Larrard, 2013] to model the limit order books with $X_k=\{-\delta,+\delta\},$ where $\delta$ is the fixed tick size.

\subsection{Regime-switching Compound Hawkes Process \\(RSCHP)}

Let $Y_t$ be an $N$-state Markov chain, with rate matrix $A_t.$ We assume, without loss of generality, that $Y_t$ takes values in the standard basis vectors in $R^N.$ Then, $Y_t$ has the representation
$$
Y_t=Y_0+\int_0^tA_sY_sds+M_t,
\eqno{(12)}
$$
for $M_t$ an $R^N$ -valued $P$-martingale (see [Buffington and Elliott, 2000] for more details).

{\bf Definition 6 (One-dimensional Regime-switching Hawkes Process)}. A one-dimensional regime-switching Hawkes Process $N_t$ is a point process characterized by its intensity $\lambda(t)$ in the following way:
$$
\lambda_t=<\lambda,Y_t>+\int_{0}^t<\mu(t-s),Y_s>dN_s,
\eqno{(13)}
$$
where $<\cdot,\cdot>$ is an inner product and $Y_t$ is defined in (12).

{\bf Definition 7 (Regime-switching Compound Hawkes Process (RSHP)).}

Let $N_t$ be any one-dimensional regime-switching Hawkes process as defined in (13), Definition 6. Let also $X_n$ be an ergodic continuous-time finite state Markov chain, independent of $N_t,$ with space state $X.$ The regime-switching compound Hawkes process is defined as
$$
S_t=S_0+\sum_{i=1}^{N_t}X_k,
\eqno{(14)}
$$
where $N_t$ is defined in (13).

{\bf Remark 3.} In similar way, as in Definition 6, we can define regime-switching Hawkes processes with exponential kernel, (see (4)), or power law kernel (see (6)).

{\bf Remark 4.} Regime-switching Hawkes processes were considered in [Cohen and Elliott, 2014] (with exponential kernel) and in [Vinkovskaya, 2014], (multi-dimensional Hawkes process).
Paper [Cohen and Elliott, 2014] discussed a self-exciting counting process whose parameters depend on a hidden finite-state Markov chain, and the optimal filter and smoother based on observations of the jump process are obtained. Thesis [Vinkovskaya, 2014] considers a regime-switching multi-dimensional Hawkes process with an exponential kernel which reflects changes in the bid-ask spread.
The statistical properties, such as maximum likelihood estimations of its parameters, etc., of this model were studied.

%%%%%%%%%%%%%%%%%%%%%%%%%%%%%%%%%%%
\section{Diffusion Limits  and LLNs for CHP and RSCHP in Limit Order Books}

In this section, we consider LLNs and diffusion limits for the CHP and RSCHP, defined above, as used in the limit order books. In the limit order books, high-frequency and algorithmic trading, order arrivals and cancellations are very frequent and occur at the millisecond time scale (see, e.g., [Cont and Larrard, 2013], [Cartea {\it et al.}, 2015]). Meanwhile, in many applications, such as order execution, one is interested in the dynamics of order flow over a large time scale, typically tens of seconds or minutes. It means that we can use asymptotic methods to study the link between price volatility and order flow in our model by studying the diffusion limit of the price process. Here, we prove functional central limit theorems for the price processes and express the volatilities of price changes in terms of parameters describing the arrival rates and price changes. In this section, we consider diffusion limits and LLNs for both CHP, sec. 3.1, and RSCHP, sec. 3.2, in the limit order books. We note, that level-1 limit order books with time dependent arrival rates $\lambda(t)$ were studied in [Chavez-Casillas {\it et al.}, 2016], including the asymptotic distribution of the price process.

\subsection{Diffusion Limits for CHP in Limit Order Books}

We consider here the mid-price process $S_t$ (CHP) which was defined in (10), as,
$$
S_t=S_0+\sum_{k=1}^{N(t)}X_k.
\eqno{(15)}
$$
Here, $X_k\in \{-\delta,+\delta\}$ is continuous-time two-state Markov chain,  $\delta$ is the fixed tick size, and $N(t)$ is the number of price changes up to moment $t,$ described by the one-dimensional Hawkes process defined in (2), Definition 4.  It means that we have the case with a fixed tick,  a two-valued price change and dependent orders.

%%%%%%%%%%%%%%%%%%%%%
{\bf Theorem 1 (Diffusion Limit for CHP).} Let $X_k$ be an ergodic Markov chain with two states $\{-\delta,+\delta\}$ and with ergodic probabilities $(\pi^*,1-\pi^*).$ Let also $S_t$ be defined in (15). Then
$$
\frac{S_{nt}-N(nt)s^*}{\sqrt{n}}\to_{n\to+\infty}\sigma\sqrt{\lambda/(1-\hat\mu)}W(t),
\eqno{(16)}
$$
where $W(t)$ is a standard Wiener process, $\hat\mu$ is given by
$$
0<\hat\mu:=\int_0^{+\infty}\mu(s)ds<1\quad and\quad \int_0^{+\infty}\mu(s)sds<+\infty,
\eqno{(17)}
$$
$$
s^*:=\delta(2\pi^*-1)\quad and\quad \sigma^2:=4\delta^2\Big(\frac{1-p'+\pi^*(p'-p)}{(p+p'-2)^2}-\pi^*(1-\pi^*)\Big).
\eqno{(18)}
$$
Here, $(p,p')$ are the transition probabilities of the Markov chain $X_k.$ We note that $\lambda$ and $\mu(t)$ are defined in (2).

{\bf Proof.} From (15) it follows that
$$
S_{nt}=S_0+\sum_{k=1}^{N(nt)}X_k,
\eqno{(19)}
$$
and
$$
S_{nt}=S_0+\sum_{k=1}^{N(nt)}(X_k-s^*)+N(nt)s^*.
$$
Therefore,
$$
\frac{S_{nt}-N(nt)s^*}{\sqrt{n}}=\frac{S_0+\sum_{k=1}^{N(nt)}(X_k-s^*)}{\sqrt{n}}.
\eqno{(20)}
$$
Since $\frac{S_0}{\sqrt{n}}\to_{n\to+\infty}0,$ we have to find the limit for
$$\frac{\sum_{k=1}^{N(nt)}(X_k-s^*)}{\sqrt{n}}$$
when $n\to+\infty.$

Consider the following sums
$$
R_n:=\sum_{k=1}^{n}(X_k-s^*)
\eqno{(21)}
$$
and
$$
U_n(t):=n^{-1/2}[(1-(nt-\lfloor nt\rfloor))R_{\lfloor nt\rfloor}+(nt-\lfloor nt\rfloor))R_{\lfloor nt\rfloor)+1}],
\eqno{(22)}
$$
where $\lfloor\cdot\rfloor$ is the floor function.

Following the martingale method from [Swishchuk and Vadori, 2015], we have the following weak convergence in the Skorokhod topology (see [Skorokhod, 1965]):
$$
U_n(t)\to_{n\to+\infty}\sigma \cW_t,
\eqno{(23)}
$$
where $\sigma$ is defined in (18), and $\cW_t$ is a standard Brownian motion.

We note that w.r.t LLN for Hawkes process $N(t)$ (see, e.g., [Daley and Vee-Jones, 2010]) we have:
$$
\frac{N(t)}{t}\to_{t\to+\infty}\frac{\lambda}{1-\hat\mu}:=\bar\lambda,
$$
or
$$
\frac{N(nt)}{n}\to_{n\to+\infty}\frac{t\lambda}{1-\hat\mu}=\bar\lambda t,
\eqno{(24)}
$$
where $\hat\mu$ is defined in (17).

Using a change of time in (23), $t\to N(nt)/n,$ we can find from (23) and (24):
$$
U_n(N(nt)/n)\to_{n\to+\infty}\sigma \cW\Big(t\lambda/(1-\hat\mu)\Big),
$$
or
$$
U_n(N(nt)/n)\to_{n\to+\infty}\sigma\sqrt{\lambda/(1-\hat\mu)}W(t),
\eqno{(25)}
$$
where $W_t = \cW_{\bar\lambda t}/\sqrt{\bar\lambda}.$ The Brownian motion $W(t)$ in (25) is equivalent by distribution to Brownian motion $\cW$ in (23) by scaling property.
The result (16) now follows from (20)-(25).

{\bf Remark 5.} In the case of exponential decay, $\mu(t)=\alpha e^{-\beta t}$ (see (4)), the limit in (16) is $[\sigma/\sqrt{\lambda/(1-\alpha/\beta)}]W(t),$ because $\hat\mu=\int_0^{+\infty}\alpha e^{-\beta s}ds=\alpha/\beta.$

%%%%%%%%%%%%%%%%
\subsection{LLN for CHP}

{\bf Lemma 1 (LLN for CHP)}. The process $S_{nt}$ in (19) satisfies the following weak convergence in the Skorokhod topology (see [Skorokhod, 1965]):
$$
\frac{S_{nt}}{n}\to_{n\to+\infty}s^*\frac{\lambda}{1-\hat\mu}t,
\eqno{(26)}
$$
where $s^*$ and $\hat\mu$ are defined in (18) and (17), respectively.

{\bf Proof.} From (19) we have
$$
S_{nt}/n=S_0/n+\sum_{k=1}^{N(nt)}X_k/n.
\eqno{(27)}
$$
The first term goes to zero when $n\to+\infty.$
From the other side, using the strong LLN for Markov chains (see, e.g., [Norris, 1997])
$$
\frac{1}{n}\sum_{k=1}^{n}X_k\to_{n\to+\infty} s^*,
\eqno{(28)}
$$
where $s^*$ is defined in (18).

Finally, taking into account (24) and (28), we obtain:
$$
\sum_{k=1}^{N(nt)}X_k/n=\frac{N(nt)}{n}\frac{1}{N(nt)}\sum_{k=1}^{N(nt)}X_k\to_{n\to+\infty}s^*\frac{\lambda}{1-\hat\mu}t,
$$
and the result in (26) follows.

{\bf Remark 6.} In the case of exponential decay, $\mu(t)=\alpha e^{-\beta t}$ (see (4)), the limit in (26) is $s^*t(\lambda/(1-\alpha/\beta)),$ because $\hat\mu=\int_0^{+\infty}\alpha e^{-\beta s}ds=\alpha/\beta.$

%%%%%%%%%%%%%%%%%%%%%%%%%%%%%%%%%%%%%%%%

\subsection{Corollary: Extension to a Point Process}

The price process $S$ is expressed as
$$
S_t = S_0 + \sum_{i=1}^{N(t)}X_i, \qquad t\ge 0,
$$
where $N$ is a point process, and Markov chain $X_i$ is defined in (10).

\textbf{Assumption C1:}
 As $n\to\infty$, $N(n t)/n \stackrel{Pr}{\longrightarrow} \bar \lambda t,$ where $\bar \lambda:=\lambda/(1-\hat\mu).$

Note that if  $N(t) = \max\{n: V_n\le t\}$, then $N(nt)/n \stackrel{Pr}{\longrightarrow} \bar \lambda t=\frac{1}{\bar v} $ iff $ V_n/n \stackrel{Pr}{\longrightarrow}\bar v$. This representation is useful in particular for renewal processes where $V_n = \sum_{k=1}^n \tau_k$, with the $\tau_k$ i.i.d. with mean $\bar v$.

\textbf{Assumption C2:} $ U_n(t) \rightsquigarrow W$, where $W$ is a Brownian motion, and $U_n(t)$ is defined in (22).

It then follows from Assumptions C1 and C2 that
$$
n^{-1/2} \left\{S_{nt}-S_0-s^*N(nt)\} \right\} = \sigma U_n\left(N(nt)/n\right) = n^{-1/2} \sum_{i=1}^{N(nt)}\{X_i - s^*\}  \rightsquigarrow  \sigma \sqrt{\bar\lambda} \; W_t ,
$$
where $W$ is a Brownian motion, and $s^*$ is denied in (18). In fact, for any $t\ge 0$,  $W_t = \cW_{\bar\lambda t}/\sqrt{\bar\lambda}$.

The limiting variance $\sigma^2 \bar\lambda$ can probably be approximated by summing the square of the increments $S_{nt_i}-S_{nt_{i-1}}-s^*(N(nt_i)-N(nt_{i-1}))$. In any cases, $\bar \lambda$ cab be easily estimated by $N(T)/T$, and $\sigma$ can be estimated from the distribution of the price increments.\\

Suppose now that there is also a CLT for the point process $N$. More precisely,

\textbf{Assumption C3:} $n^{1/2}\left( \frac{N(nt)}{n}-t\lambda\right)    \rightsquigarrow \bar\sigma \bar W_t$, where $\bar W$ is a Brownian motion independent of $W.$

Then under Assumptions C1--C3,

%\begin{equation}\label{eq:Sdecomp}
$$
n^{-/2}\left\{S_{n t} - n  t \bar\lambda s^*\right\} \rightsquigarrow \tilde\sigma  \dW_t,
$$
%\end{equation}

where $\dW  =  \left\{\sigma\sqrt{\bar\lambda} W+ s^* \bar\sigma\bar W\right\}/\tilde\sigma  $ is a Brownian motion, and

%\begin{equation}\label{eq:sigmatilde}
$$
\tilde \sigma = \left[ \sigma^2 \bar\lambda + \{s^*\}^2 \bar\sigma^2 \right]^{1/2}.
$$
%\end{equation}

This follows from Assumptions and the fact that
$$
n^{-1/2}\left\{S_{n t} - S_0-  n  t \bar\lambda s^*\right\} = n^{-1/2}\sum_{i=1}^{N(n t)}\{X_i -s^*\} + s^* n^{1/2}\left( \frac{N(nt)}{n}-t\lambda\right) .
$$

%\begin{rem}
{\bf Remark 7}.
Assumption C3 is true in many interesting cases. For renewal processes, if $\sigma_\tau$ is the standard deviation of $\tau_k$, then $\bar\sigma =  \sigma_\tau \bar\lambda^{3/2} $. This is also  true for Hawkes processes [Bacry {\it et al.}, 2013] with $\lambda(t) = \lambda_0+\int_0^t \mu(t-s)dN_s$, provided $\hat\mu = \int_0^\infty \mu(s)ds <1$. Then $\bar \lambda = \frac{\lambda}{1-\hat\mu}$ and
$\bar \sigma = \sqrt{\bar\lambda} /(1-\hat\mu)$.

%One could also model several prices by using a multivariate Hawkes process, as defined in \cite{BDHM}.
%\end{rem}

%%%%%%%%%%%%%%%%%%%%%%%%%%%%%%%%%%%%%%%%

%%%%%%%%%%%%%%%%%%%%%%
\subsection{Diffusion Limits for RSCHP in Limit Order Books}

Consider now the mid-price process $S_t$ (RSCHP) in the form
$$
S_t=S_0+\sum_{k=1}^{N_t}X_k,
\eqno{(29)}
$$
where $X_k\in \{-\delta,+\delta\}$ is continuous-time two-state Markov chain,  $\delta$ is the fixed tick size, and $N_t$ is the number of price changes up to the moment $t,$ described by a one-dimensional regime-switching Hawkes process with intensity given by:
$$
\lambda_t=<\lambda,Y_t>+\int_{0}^t\mu(t-s)dN_s,
\eqno{(30)}
$$
(compare with (11), Definition 6).

Here we would like to relax the model for one-dimensional regime-switching Hawkes process, considering only the case of a switching the parameter $\lambda,$ background intensity, in (20), which is reasonable from a limit order book's point of view. For example, we can consider a three-state Markov chain $Y_t\in\{e_1,e_2,e_3\}$ and interpret $<\lambda,Y_t>$ as the imbalance, where $\lambda_1, \lambda_2,  \lambda_3,$ represent high, normal and low imbalance, respectively (see [Cartea {\it et al.}, 2015] for imbalance notion and discussion). Of course, a more general case (13) can be considered as well, where the excitation function $<\mu(t),Y_t>,$, can take three values, corresponding to high imbalance, normal imbalance, and low imbalance, respectively.

%%%%%%%%%%%%%%%%%%%%%%%
{\bf Theorem 2 (Diffusion Limit for RSCHP)}. Let $X_k$ be an ergodic Markov chain with two states $\{-\delta,+\delta\}$ and with ergodic probabilities $(\pi^*,1-\pi^*).$ Let also $S_t$ be defined in (29) with $\lambda_t$ as in (30). We also consider $Y_t$ to be an ergodic Markov chain with ergodic probabilities $(p_1^*,p_2^*,...,p_N^*).$ Then
$$
\frac{S_{nt}-N_{nt}s^*}{\sqrt{n}}\to_{n\to+\infty}\sigma\sqrt{\hat\lambda/(1-\hat\mu)}W(t),
\eqno{(31)}
$$
where $W(t)$ is a standard Wiener process with $s^*$ and $\sigma$ defined in (18),
$$
\hat\lambda:=\sum_{i=1}^Np_i^*\lambda_i\not=0,\quad \lambda_i:=<\lambda,i>,
\eqno{(32)}
$$
and $\hat\mu$ is defined in (17).

{\bf Proof.} From (29) it follows that
$$
S_{nt}=S_0+\sum_{i=1}^{N_{nt}}X_k,
\eqno{(33)}
$$
and
$$
S_{nt}=S_0+\sum_{i=1}^{N_{nt}}(X_k-s^*)+N_{nt}s^*,
$$
where $N_{nt}$ is an RGCHP with regime-switching intensity $\lambda_t$ as in (30).
Then,
$$
\frac{S_{nt}-N_{nt}s^*}{\sqrt{n}}=\frac{S_0+\sum_{i=1}^{N_{nt}}(X_k-s^*)}{\sqrt{n}}.
\eqno{(34)}
$$
As long as $\frac{S_0}{\sqrt{n}}\to_{n\to+\infty}0,$ we wish to find the limit of
$$\frac{\sum_{i=1}^{N_{nt}}(X_k-s^*)}{\sqrt{n}}$$
when $n\to+\infty.$

Consider the following sums, similar to (21) and (22):
$$
R_n:=\sum_{k=1}^{n}(X_k-s^*)
\eqno{(35)}
$$
and
$$
U_n(t):=n^{-1/2}[(1-(nt-\lfloor nt\rfloor))R_{\lfloor nt\rfloor)}+(nt-\lfloor nt\rfloor))R_{\lfloor nt\rfloor)+1}],
\eqno{(36)}
$$
where $\lfloor\cdot\rfloor$ is the floor function.

Following the martingale method from [Swishchuk and Vadori, 2015], we have the following weak convergence in the Skorokhod topology (see [Skorokhod, 1965]):
$$
U_n(t)\to_{n\to+\infty}\sigma W(t),
\eqno{(37)}
$$
where $\sigma$ is defined in (18).

We note that with respect to the LLN for the Hawkes process $N_t$ in (34)  with regime-switching intensity $\lambda_t$ as in (30) we have (see [Korolyuk and Swishchuk, 1995] for more details):
$$
\frac{N_{t}}{t}\to_{t\to+\infty}\frac{\hat\lambda}{1-\hat\mu},
$$
or
$$
\frac{N_{nt}}{n}\to_{n\to+\infty}\frac{t\hat\lambda}{1-\hat\mu},
\eqno{(38)}
$$
where $\hat\mu$ is defined in (17) and $\hat\lambda$ in (32).

Using a change of time in (37), $t\to N_{nt}/n,$ we can find from (37) and (38):
$$
U_n(N_{nt}/n)\to_{n\to+\infty}\sigma W\Big(t\hat\lambda/(1-\hat\mu)\Big),
$$
or
$$
U_n(N_{nt}/n)\to_{n\to+\infty}\sigma\sqrt{\hat\lambda/(1-\hat\mu)}W(t),
\eqno{(39)}
$$
The result (31) now follows from (33)-(39).

{\bf Remark 8.} In the case of exponential decay, $\mu(t)=\alpha e^{-\beta t}$ (see (4)), the limit in (31) is $[\sigma\sqrt{\hat\lambda/(1-\alpha/\beta)}]W(t),$ because $\hat\mu=\int_0^{+\infty}\alpha e^{-\beta s}ds=\alpha/\beta.$

%%%%%%%%%%%%%%%%%%

\subsection{LLN for RSCHP}

{\bf Lemma 2 (LLN for RSCHP)}. The process $S_{nt}$ in (33) satisfies the following weak convergence in the Skorokhod topology (see [Skorokhod, 1965]):
$$
\frac{S_{nt}}{n}\to_{n\to+\infty}s^*\frac{\hat\lambda}{1-\hat\mu}t,
\eqno{(40)}
$$
where $s^*,$ $\hat\lambda$ and $\hat\mu$ are defined in (13), (27) and (12), respectively.

{\bf Proof.} From (33) we have
$$
S_{nt}/n=S_0/n+\sum_{i=1}^{N_{nt}}X_k/n,
\eqno{(41)}
$$
where $N_{nt}$ is a Hawkes process with regime-switching intensity $\lambda_t$ in (30).

The first term goes to zero when $n\to+\infty.$

From the other side, with respect to the strong LLN for Markov chains (see, e.g., [Norris, 1997])
$$
\frac{1}{n}\sum_{k=1}^{n}X_k\to_{n\to+\infty} s^*,
\eqno{(42)}
$$
where $s^*$ is defined in (18).

Finally, taking into account (38) and (42), we obtain:
$$
\sum_{i=1}^{N_{nt}}X_k/n=\frac{N_{nt}}{n}\frac{1}{N_{nt}}\sum_{i=1}^{N_{nt}}X_k\to_{n\to+\infty}s^*\frac{\hat\lambda}{1-\hat\mu}t.
$$
The result in (40) follows.

{\bf Remark 9.} In the case of exponential decay, $\mu(t)=\alpha e^{-\beta t}$ (see (4)), the limit in (40) is $s^*t(\hat\lambda/(1-\alpha/\beta)),$ because $\hat\mu=\int_0^{+\infty}\alpha e^{-\beta s}ds=\alpha/\beta.$

\section{Numerical Examples and Parameters Estimations}

%%%%%%%%%%%%%
Formula (16) in Theorem 1 (Diffusion Limit for CHP) relates the volatility of intraday returns at lower frequencies to the high-frequency arrival rates of orders. The typical time scale for order book events are milliseconds. Formula (16) states that, observed over a larger time scale, e.g., 5, 10 or 20 minutes, the price has a diffusive behaviour with a diffusion coefficient given by the coefficient at $W(t)$ in (16):
$$
\sigma\sqrt{\lambda/(1-\hat\mu)},
\eqno{(43)}
$$
where all the parameters here are defined in (17)-(18).  We mention, that this formula (43) for volatility contains all the initial parameters of the Hawkes process, Markov chain transition and stationary probabilities and the tick size. In this way, formula (43) links properties of the price to the properties of the order flow. 

Also, the left hand side of (16) represents the variance of price changes, whereas the right hand side in (16) only involves the tick size and Hawkes process and Markov chain quantities. From here it follows that an estimator for price volatility may be computed without observing the price at all.
As we shall see below, the error of estimation of comparison of the standard deviation of the LNS of (16) and the RHS of (16) multiplied by $\sqrt{n}$  is approximately 0.08, indicating that approximation in (16) for diffusion limit for CHP in Theorem 1, is pretty good.

Section 4.1 below presents parameters estimation for our model using CISCO Data (5 Days, 3-7 Nov 2014 (see [Cartea {\it et al.}, 2015])). Section 4.2 contains the errors of estimation of comparison of of the standard deviation of the LNS of (16) and the RHS of (16) multiplied by $\sqrt{n}.$ Section 4.3 depicts some graphs based on parameters estimation from sec. 4.1. And Section 4.4  presents some ideas of how to implement the regime switching case from sec. 3.4.

%%%%%%%%%%%%%

\subsection{Parameters Estimation for CISCO Data (5 Days, 3-7 Nov 2014 (see [Cartea {\it et al.}, 2015]))}

We have the following estimated parameters for 5 days, 3-7 November 2014, from Formula (16):

$$
\begin{array}{rcl}
\vspace{0.5cm}

s^*&=&
0.0001040723; 0.0002371220; 0.0002965143; 0.0001263690; 0.0001554404;\\

\vspace{0.5cm}

\sigma&=&
1.066708e-04; 1.005524e-04; 1.165201e-04; 1.134621e-04; \\&&9.954487e-05;\\

\vspace{0.5cm}

\lambda&=&
0.03238898 ;
0.02643083 ;
0.02590728 ;
0.02530517 ;
0.02417804 ;\\

\vspace{0.5cm}

\alpha&=&
438.2557; 401.0505;
559.1927; 418.7816; 449.8632;\\

\vspace{0.5cm}

\beta&=&
865.9344; 718.0325; 1132.0741; 834.2553; 878.9675;\\

\hat\lambda:=\lambda/(1-\alpha/\beta)&=&
0.06560129;
0.059801686;
0.051181133;
0.050801432;
0.04957073.\\
\end{array}
$$

Volatility Coefficient $\sigma\sqrt{\lambda/(1-\alpha/\beta)}$ (volatility coefficient for the Brownian Motion in the right hand-side (RHS) of (16)):

$$
0.04033114;
0.04098132;
0.04770726;
0.04725449;
0.04483260.\\
$$
\vspace{0.5cm}
Transition Probabilities $p:$

\vspace{0.5cm}

Day1:
$$
\begin{array}{rcl}
                    uu         &&            ud\\
               0.5187097  &&    0.4812903\\
                du            &&          dd\\
               0.4914135  &&    0.5085865\\
  \end{array}
 $$

Day2:
$$
\begin{array}{rcl}
0.4790503  &&  0.5209497 \\
0.5462555   && 0.4537445  \\
\end{array}
$$

Day3:
$$
\begin{array}{rcl}
0.6175041       &&      0.3824959\\
0.4058722        &&      0.5941278\\
\end{array}
$$

Day4:
$$
\begin{array}{rcl}
0.5806988     &&         0.4193012\\
0.4300341     &&         0.5699659\\
\end{array}
$$

Day5:
$$
\begin{array}{rcl}
0.4608844     &&      0.5391156\\
0.5561404    &&      0.4438596\\
\end{array}
$$

We note, that stationary probabilities $\pi_i^*, i=1,...,5,$ are, respectively: $0.5525; 0.6195; 0.6494; 0.5637; 0.5783.$ Here, we assume that the tick $\delta$ size is $\delta=0.01.$

The following set of parameters are related to the the following expression $$S_{nt}-N(nt)s^*=S_0+\sum_{k=1}^{N(nt)}(X_k-s^*),$$-LHS of the expression in (16) multiplied by $\sqrt{n}.$

The first set of numbers are for the 10 minutes time horizon ($nt=10$ minutes, for 5 days, the 7 sampled hours, total 35 numbers):

\begin{center}
Table 1
\end{center}
$$
[1] 24.50981;
[2] 24.54490;
[3] 24.52375;
[4] 24.59209;
[5] 24.47209;
[6] 24.57042;
[7] 24.61063;
 $$
 $$
[8] 24.76987;
[9] 24.68749;
[10] 24.81599;
[11] 24.77026;
[12] 24.79883;
[13] 24.80073;
[14] 24.90121;
 $$
 $$
[15] 24.87772;
[16] 24.98492;
[17] 25.09788;
[18] 25.09441;
[19] 24.99085;
[20] 25.18195;
[21] 25.15721;
 $$
 $$
[22] 25.04236;
[23] 25.18323;
[24] 25.15222;
[25] 25.20424;
[26] 25.14171;
[27] 25.18323;
[28] 25.25348;
 $$
 $$
[29] 25.10225;
[30] 25.29003;
[31] 25.28282;
[32] 25.33267;
[33] 25.30313;
[34] 25.27407;
[35] 25.30438;
$$

The standard deviation (SD) is:  $ 0.2763377.$ The Standard Error (SE) for SD for the 10 min is: $0.01133634$ (for standard error calculations see [Casella and Berger, 2002, page 257].

The second set of numbers are for the 5 minutes time horizon ($nt=5$ minutes, for 5 days, the 7 sampled hours):
\begin{center}
Table 2
\end{center}$$
[1] 24.49896;
[2] 24.52906;
[3] 24.50417;
[4] 24.53417;
[5] 24.53500;
[6] 24.51458;
[7] 24.55479;
$$
$$
[8] 24.93026;
[9] 24.66931;
[10] 24.74263;
[11] 24.79358;
[12] 24.80310;
[13] 24.84500;
[14] 24.88405;
$$
$$
[15] 24.85729;
[16] 24.98907;
[17] 25.08085;
[18] 25.07500;
[19] 24.99322;
[20] 25.13381;
[21] 25.15144;
$$
$$
[22] 25.15197;
[23] 25.12475;
[24] 25.15449;
[25] 25.18475;
[26] 25.20348;
[27] 25.20500;
[28] 25.25348;
$$
$$
[29] 25.21251;
[30] 25.35376;
[31] 25.30407;
[32] 25.30469;
[33] 25.30469;
[34] 25.27500;
[35] 25.30469;
 $$

The standard deviation for those numbers is: $0.2863928.$ The SE for SD for the 5 min is: $0.01233352.$

The third and last set of numbers are for the 20 minutes time horizon ($nt=20$ minutes, for 5 days, the 7 sampled hours):
\begin{center}
Table 3
\end{center} $$
 [1] 24.48419;
[2] 24.53970;
[3] 24.56292;
[4] 24.57105;
[5] 24.48938;
[6] 24.52751;
[7] 24.50751;
$$
$$
[8] 24.76465;
[9] 24.59753;
[10] 24.82935;
[11] 24.76552;
[12] 24.81741;
[13] 24.75409;
[14] 24.84077;
$$
$$
[15] 24.92942;
[16] 24.99721;
[17] 25.05551;
[18] 25.04848;
[19] 25.08492;
[20] 25.09780;
[21] 25.09551;
$$
$$
[22] 24.95124;
[23] 25.24222;
[24] 25.19096;
[25] 25.18273;
[26] 25.14070;
[27] 25.20171;
[28] 25.26785;
$$
$$
[29] 25.23013;
[30] 25.38661;
[31] 25.32127;
[32] 25.34065;
[33] 25.30313;
[34] 25.25251;
[35] 25.24972;
 $$

The standard deviation is: $0.2912967.$ The SE for SD for the 20 min is: $0.01234808.$

 As we can see, the SE is approximately $0.01$ for all three cases.

%%%%%%%%%%%%%%%%%%%%%%%%%%%

\subsection{Error of Estimation}

Here, we would like to calculate the error of estimation comparing the standard deviation for $$S_{nt}-N(nt)s^*=S_0+\sum_{k=1}^{N(nt)}(X_k-s^*)$$ and standard deviation in the right-hand side of (16) multiplied by $\sqrt{n},$ namely, $$\sqrt{n}\sigma\sqrt{\lambda/(1-\alpha/\beta)}.$$

We calculate the error of estimation with respect to the following formula:
$$
ERROR = (1/m) \sum_{k=1}^m (sd-\hat{sd})^2,
$$
where $\hat{sd} = \sqrt{n}Coef,$ where $Coef$ is the volatility coefficient in the right-hand side of equation (16). In this case $n=1000,$ and $Coef =0.3276.$

We take observations of $S_{nt}-N(tn)s^*$ every 10 min and we have 36 samples per day for 5 days.

Using the first approach with  formula above we take $m=5$ and for computing the standard deviation "sd" we take 36 samples of the first day.  In that case, we have

$ERROR = 0.07617229.$

Using the second approach with
formula above, we take $m=36$ and for computing "sd" we take samples of 5 elements (the same time across 5 days). In that case we have

$ERROR= 0.07980041.$

As we can see, the error of estimation in both cases is approximately $0.08,$ indicating that approximation in (16) for diffusion limit for CHP, Theorem 1, is pretty good.
%%%%%%%%%%%%%%%%%%%%%%%%%%%%%%%

\subsection{Graphs based on Parameters Estimation for CISCO Data (5 Days, 3-7 Nov 2014 ([Cartea {\it et al.}, 2015])) from Sec. 4.1}
%%%%%%%%%%%%%%%%%%%%%%%%

%%%%%%%%%%%%%%%%%%%%%%%%

The following graphs contain the empirical intensity for the point process for those 5 days vs a simulated path using the above-estimated parameters.

\begin{minipage}[t]{1.25\linewidth}
        \centering
        \includegraphics[width=\columnwidth]{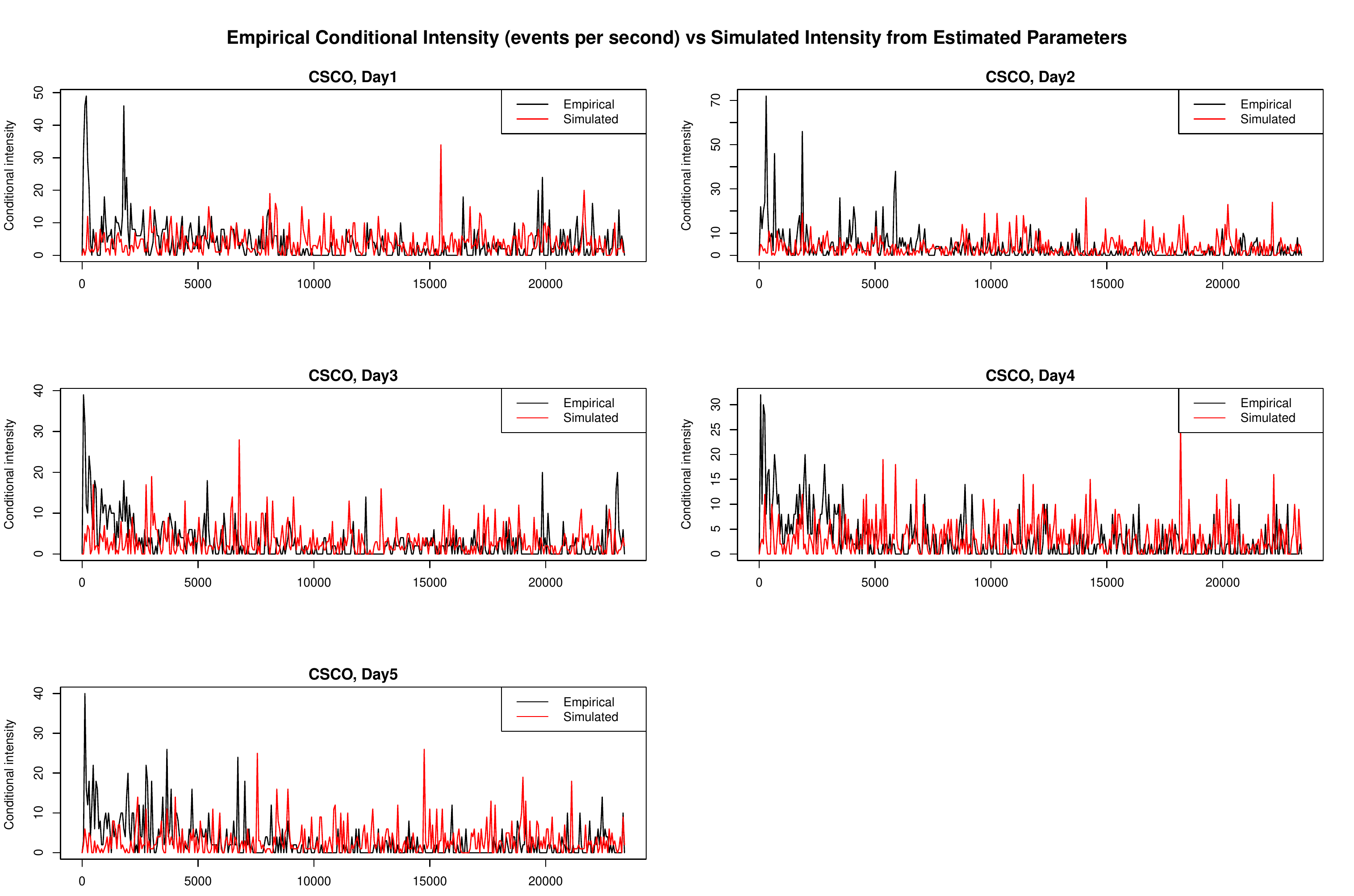}
        {\small }
         \end{minipage}

%%%%%%%%%%%%%%%%
In the next graphs we estimate the left hand-side (LHS) of (16). The time horizon is $nt=10$ min. We took the time from which the start time measuring the 10 min. as the independent variable or $x$-axis.  The dependent variable or $y$-axis is
$$F(t_0)=  ( S_{t_0} + S_{tn}  - N(tn)s^* )/\sqrt{n}.$$

\begin{minipage}[t]{1.25\linewidth}
        \centering
        \includegraphics[width=\columnwidth]{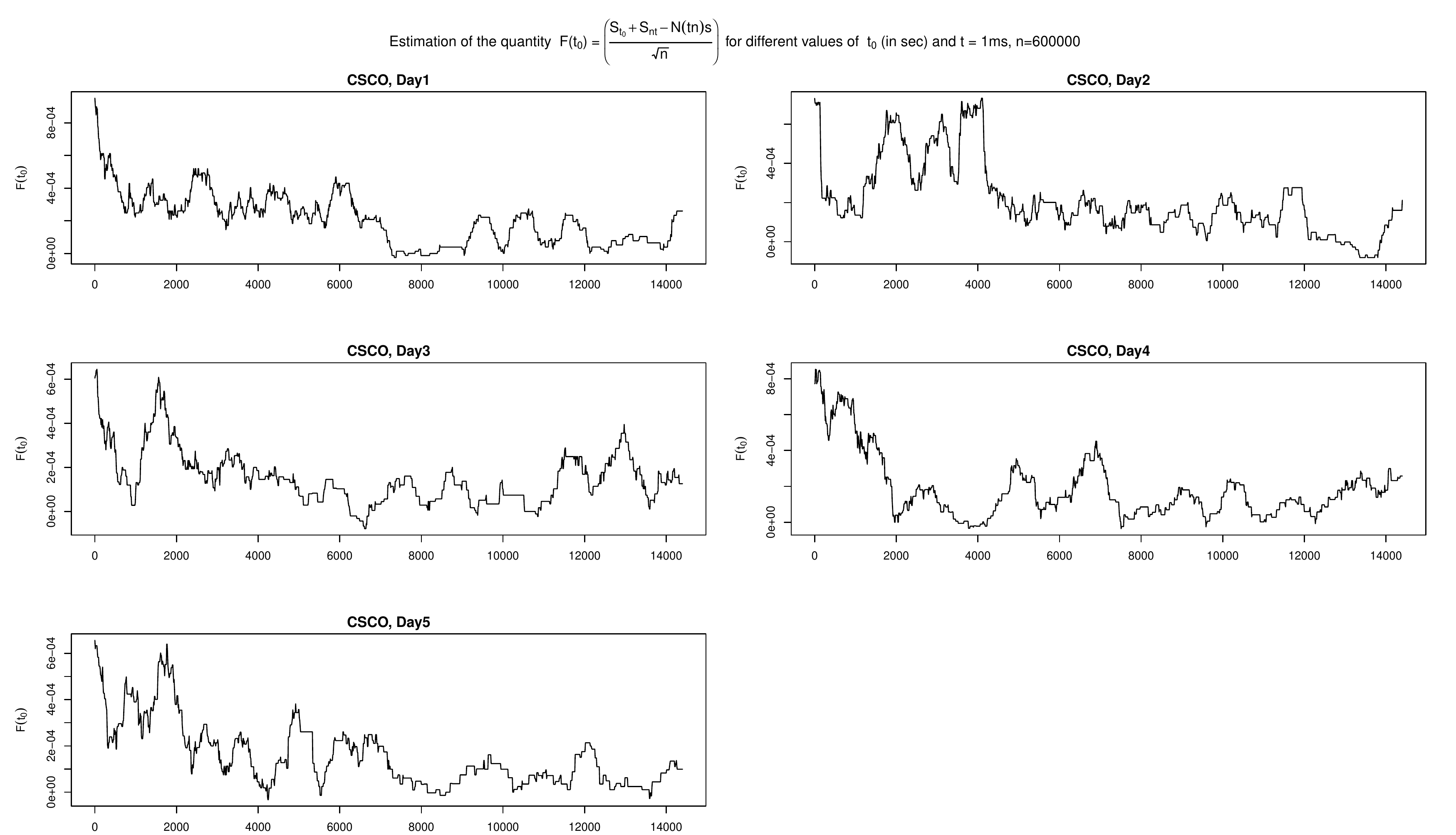}
        {\small }
         \end{minipage}

The following graphs are the same as above but just considering the median of the $1000$ simulations and zoomed in the range so that it is easy to compare.

\begin{minipage}[t]{1.25\linewidth}
        \centering
        \includegraphics[width=\columnwidth]{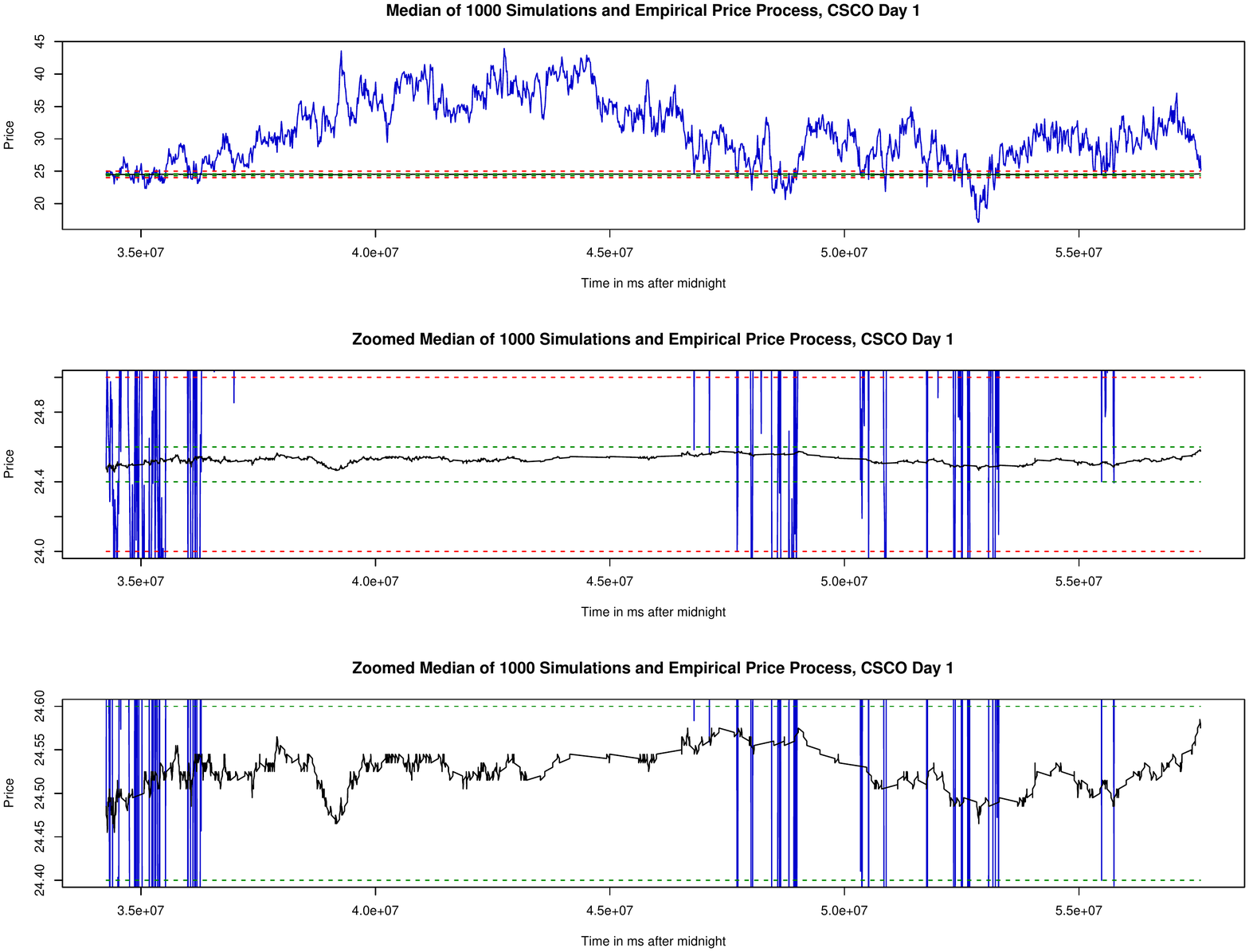}
        {\small }
         \end{minipage}

The next graphs contain information on the quantiles of simulations of the price process according to equation (16). That is, for a fixed big $n$  and fixed $t_0$ and $t.$ We use $1000$ simulations of the process (with the parameters estimated for $N(t)$). The time horizon is a trading day.

\begin{minipage}[t]{1.1\linewidth}
        \centering
        \includegraphics[width=\columnwidth]{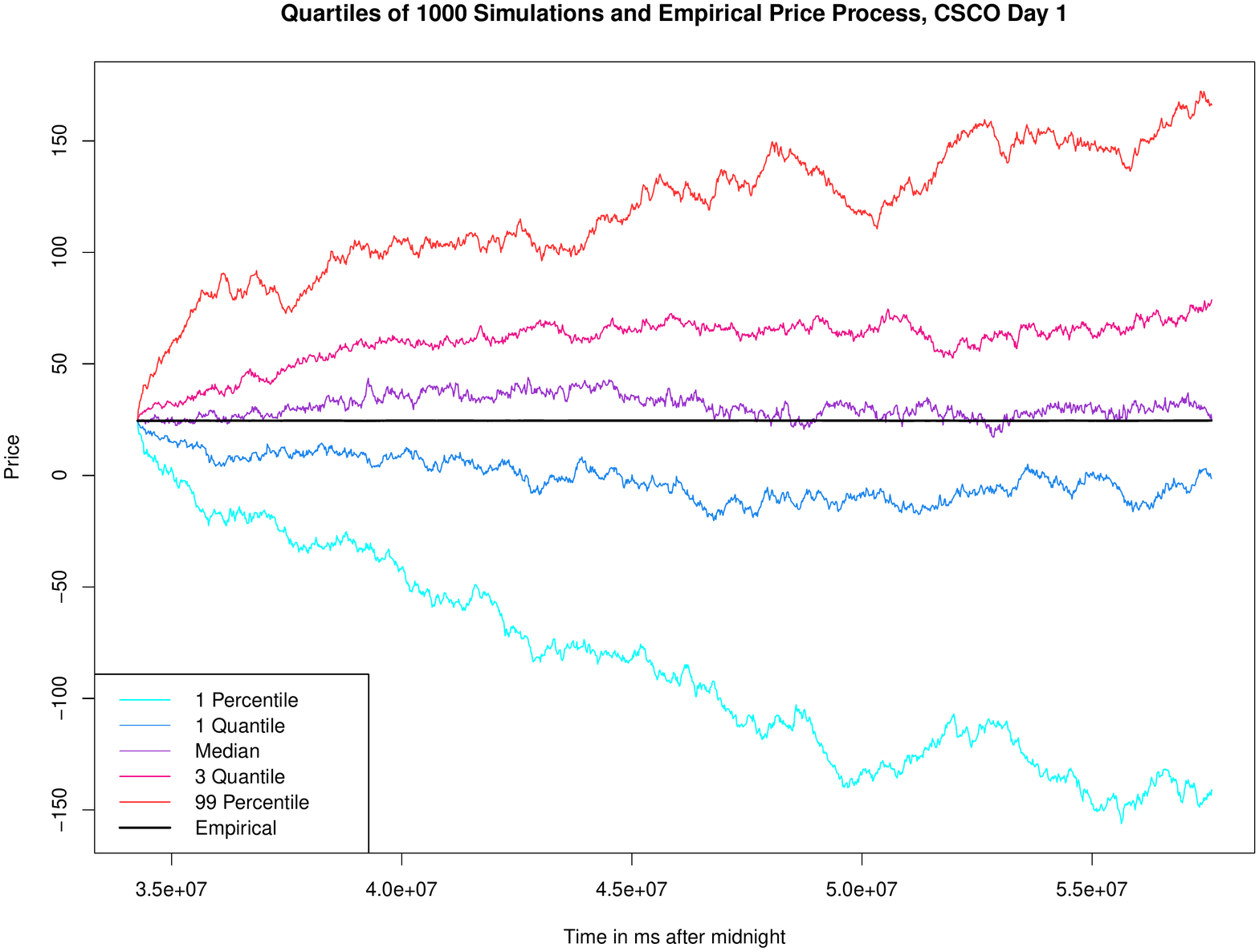}
        {\small }
         \end{minipage}

The following graph is the same as above but the time horizon is $5$ minutes (e.g., $nt=5$ minutes now, $n$ is the same).

\begin{minipage}[t]{1.1\linewidth}
        \centering
        \includegraphics[width=\columnwidth]{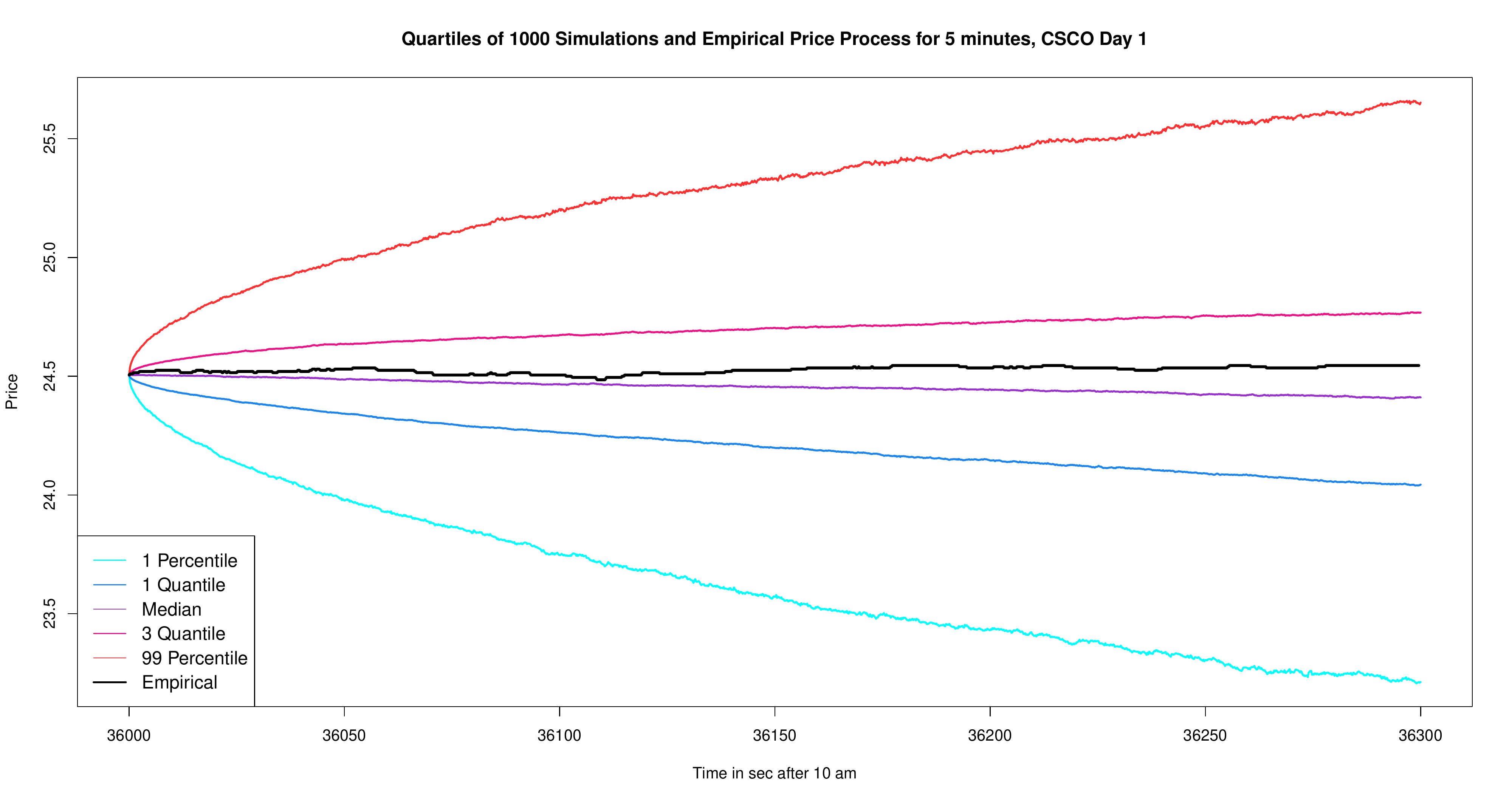}
        {\small }
         \end{minipage}

The last graph is the same as above but the time horizon is $60$ minutes (e.g., $nt=60$ minutes now, $n$ is the same).

\begin{minipage}[t]{1.1\linewidth}
        \centering
        \includegraphics[width=\columnwidth]{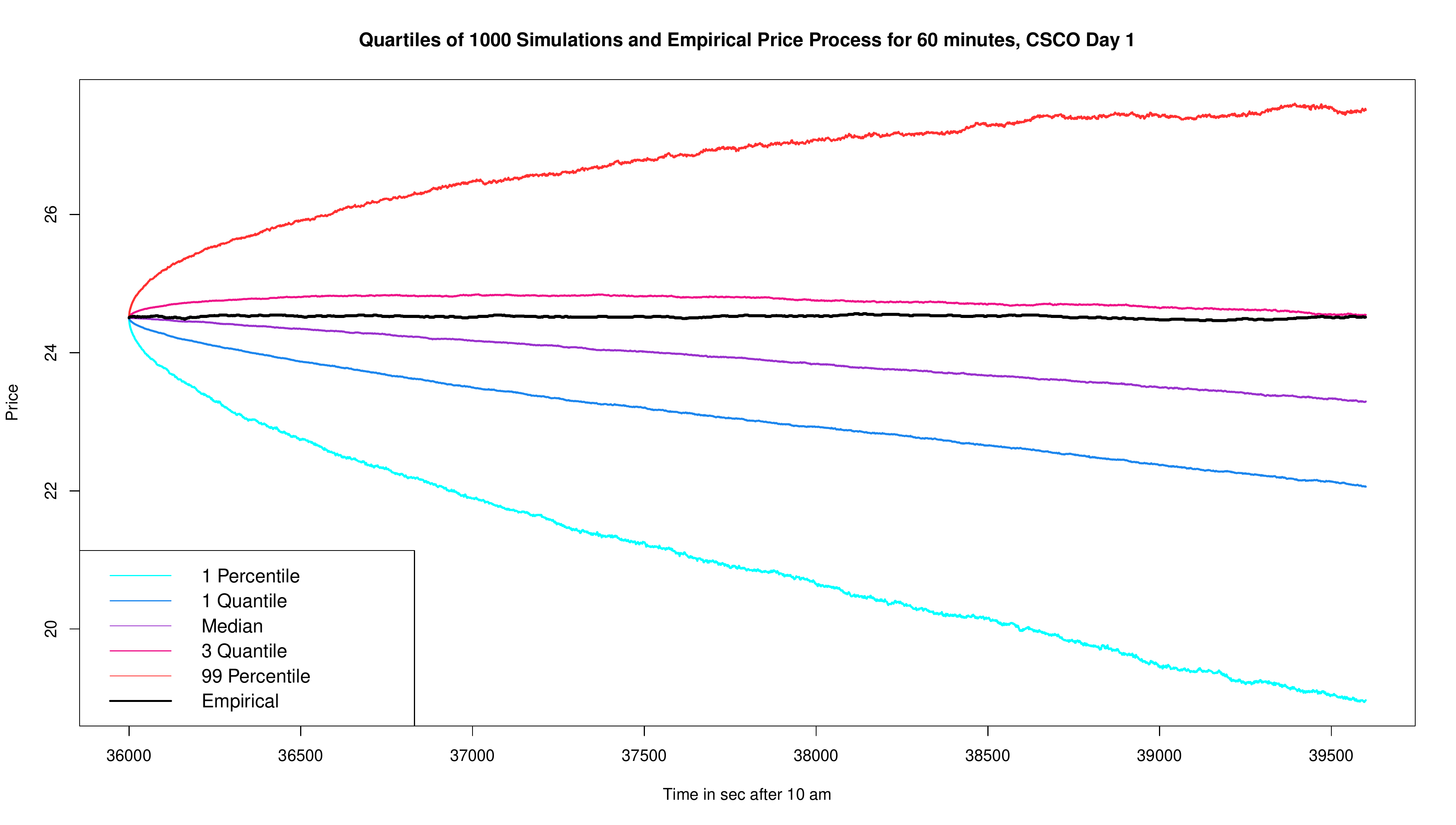}
        {\small }
         \end{minipage}

\subsection{Remark on Regime-switching Case (sec. 3.4)}

We present here some ideas of how to implement the regime-switching case from section 3.4. We take a look at the case of two states for intensity $\lambda.$ The first state is constructed as the intensity that is above the intensities average, and the second state is constructed as the intensity that is below the intensities average. The transition probabilities matrix $P$ are calculated using the relative frequencies of the intensities, and the stationary probabilities $\vec p=(p_1,p_2)$ are calculated from the equation $\vec p P=\vec p.$ Then $\hat\lambda$ can be calvculated from formula (32). For example, for the case of 5 days CISCO data we have $\lambda_1=0.03238898,$ $\lambda_2=0.02545533$ and $(p_1,p_2)=(0.2,0.8).$ In this way, the value for $\hat\lambda$ in (32) is $\hat\lambda=0.02688.$ As we could see from the data for $\lambda$ in sec. 4.1 and the latter number, the error does not exceed $0.0055.$ It means that the errors of estimation for our standard deviations in sec. 4.2 is almost the same. This is the evidence that in the case of regime-switching CHP the diffusion limit gives a very good approximation as well.

\section{Acknowledgements:} The authors wish to thank IFSID (Institut de la Finance Structur\'{e}e et des Instruments D\'{e}riv\'{e}s), Montr\'{e}al, Qu\'{e}bec, Canada, for financial support of this project. Robert Elliott also wishes to thank the SSHRC and ARC for continuing support, and  the rest of the the authors wish to thank NSERC for continuing support.

\section*{References}
\hspace{0.5cm}

Ait-Sahalia, Y., Cacho-Diaz, J. and Laeven, R. (2010). Modelling of financial contagion using mutually exciting jump processes. {\it Tech. Rep.}, 15850, Nat. Bureau of Ec. Res., USA.

Bacry, E., Delattre, S., Hoffman, M. and Muzy, J.-F. (2013). Some limit theorems for Hawkes processes and application to financial statistics. {\it Stochastic Processes and their Applications},  v. 123, No. 7, pp. 2475-2499.

Bacry, E., Mastromatteo, I. and Muzy, J.-F. (2015). Hawkes processes in finance. arXiv:1502.04592v2 [q-fin.TR] 17, May.

Bowsher, C. (2007). Modelling security market events in continuous time: intensity based, multivariate point process models. {\it J. Econometrica}, 141 (2), pp. 876-912.

Bauwens, L. and Hautsch, N. (2009). {\it Modelling Financial High Frequency Data Using Point Processes}. Springer.

Br\'{e}maud, P. and  Massouli\'{e}, L. (1996). Stability of nonlinear Hawkes processes. {\it The Annals of Probab.}, 24(3), 1563.

Buffington, J., Elliott, R.J. (2002). American Options with Regime Switching. {\it International Journal of Theoretical and Applied Finance} 5, pp. 497-514.

Buffington, J. and Elliott, R. J. (2000). Regime Switching and European Options. Lawrence, K.S. (ed.) {\it Stochastic Theory and Control}. Proceedings of a Workshop, 73-81. Berlin Heidelberg New York: Springer.

Cartea, A., Jaimungal, S. and Ricci, J. (2011). Buy low, sell high: a high-frequency trading prospective. {\it Tech. Report.}

Cartea, \'{A}., Jaimungal, S. and Penalva, J. (2015). {\it Algorithmic and High-Frequency Trading}. Cambridge University Press.

Cartensen, L. (2010). {\it Hawkes processes and combinatorial transcriptional regulation}. PhD Thesis, University of Copenhagen.

Casella, G. and Berger, R. (2002). {\it Statistical Inference}. Duxbury-Thompson Learning Inc.

Chavez-Demoulin, V. and McGill, J. (2012). High-frequency financial data modelling using Hawkes processes. {\it J. Banking and Finance}, 36(12), pp. 3415-3426.

Chavez-Casillas, J., Elliott, R., Remillard, B. and Swishchuk, A. (2017). A level-1 limit order book with time dependent arrival rates. {\it Proceed. IWAP}, Toronto, June-20-25. Also available on arXiv: https://arxiv.org/submit/1869858

Cohen, S. and Elliott, R. (2014). Filters and smoothness for self-exciting Markov modulated counting process. {\it IEEE Trans. Aut. Control}.

Cont, R. and de Larrard, A. (2013). A Markovian modelling of limit order books. {\it SIAM J. Finan. Math.}, 4(1), pp. 1-25.

Cox, D. (1955). Some statistical methods connected with series of events. {\it J. R. Stat.Soc.}, ser. B, 17 (2), pp. 129-164.

Daley, D.J. and Vere-Jones, D. (1988). {\it An Introduction to the theory of Point Processes}. Springer.

Dassios, A. and Zhao, H. (2011). A dynamic contagion process. {\it Advances in Applied Probab.}, 43(3), pp. 814-846.

Embrechts, P., Liniger, T. and Lin, L. (2011). Multivariate Hawkes processes: an application to financial data. {\it J. Appl. Prob.}, 48, A, pp. 367-378.

Errais, E., Giesecke, K. and Goldberg, L. (2010).  Affine point processes and portfolio credit risk. {\it SIAM J. Fin. Math.} 1, pp.  642-665.

Fillimonov, V., Sornette, D., Bichetti, D. and Maystre, N. (2013). Quantifying of the high level of endogeneity and of structural regime shifts in comodity markets, 2013.

Fillimonov, V. and Sornette, D. (2012). Quantifying reflexivity in financial markets: Toward a prediction of flash crashes. {\it Physical Review E}, 85(5):056108.

Hawkes, A. (1971). Spectra of some self-exciting and mutually exciting point processes. {\it Biometrica}, 58, pp. 83-90.

Hawkes, A. and Oakes, D. (1974): A cluster process representation of a self-exciting process. {\it J. Applied Probab.}, 11, pp. 493-503.

Korolyuk, V. S. and Swishchuk, A. V. (1995). Semi-Markov Random Evolutions. {\it Kluwer Academic Publishers}, Dordrecht, The Netherlands.

Laub, P., Taimre, T. and Pollett, P. (2015). Hawkes Processes.arXiv: 1507.02822v1[math.PR]10 Jul 2015.

Liniger, T. (2009). {\it Multivariate Hawkes Processes}. PhD thesis, Swiss Fed. Inst. Tech., Zurich.

McNeil, A., Frey, R. and Embrechts, P. (2015). Quantitative Risk Management: Concepts, Techniques and Tools. {\it Princeton Univ. Press}.

Mehdad, B. and Zhu, L. (2014). On the Hawkes process with different exciting functions. {\it arXiv: 1403.0994.}

Norris, J. R. (1997). Markov Chains. In Cambridge Series in Statistical and Probabilistic Mathematics. UK: Cambridge University Press.

Skorokhod, A. (1965). Studies in the Theory of Random Processes, Addison-Wesley, Reading, Mass.,
(Reprinted by Dover Publications, NY).

Swishchuk, A. (2017). Risk model based on compound Hawkes process. Abstract, IME 2017, Vienna.

 Swishchuk, A. and Vadori, N. (2017). A semi-Markovian modelling of limit order markets. {\it SIAM J. Finan. Math.}, v.8, pp. 240-273.

Swishchuk, A., Cera, K., Hofmeister, T. and Schmidt, J. (2017). General semi-Markov model for limit order books. {\it Intern. J. Theoret. Applied Finance}, v. 20, 1750019.

Swishchuk, A. and Vadori, N. (2015). Strong law of large numbers and central limit theorems for functionals of inhomogeneous Semi-Markov processes. {\it Stochastic Analysis and Applications}, 13 (2), pp. 213-243.
 
 Vinkovskaya, E. (2014). {\it A point process model for the dynamics of LOB}. PhD thesis, Columbia Univ.
 
 Zheng, B., Roueff, F. and Abergel, F. (2014). Ergodicity and scaling limit of a constrained multivariate Hawkes process. {\it SIAM J. Finan. Math.}, 5.
 
Zhu, L. (2013). Central limit theorem for nonlinear Hawkes processes, {\it J. Appl. Prob.}, 50(3), pp. 760-771.

\end{document}